\documentclass{mn2e}
\usepackage{times}
\input{psfig.sty}

\begin{document}

%\draft              

%%%%% AUTHORS -- PLACE YOUR OWN MACROS HERE %%%%%

\def\be{\begin{equation}}
\def\ee{\end{equation}}
\def\mdot{$\dot{m}$ }
\def\mpc{\,{\rm {Mpc}}}
\def\kpc{\,{\rm {kpc}}}
\def\kms{\,{\rm {km\, s^{-1}}}}
\def\msun{{$\rm M_{\odot}$}}
\def\Gyr{{\,\rm Gyr}}
\def\erg{{\rm erg}}
\def\sr{{\rm sr}}
\def\hz{{\rm Hz}}
\def\cm{{\rm cm}}
\def\sec{{\rm s}}
\def\eV{{\rm \ eV}}
\def\ledd{$L_{Edd}$~}
\def\mic{$\mu$ }
\def\ang{\AA }  % Angstrom
\def\cm2{cm$^2$ }
\def\se1{s$^{-1}$ }

\def\arcmin{\hbox{$^\prime$}}
\def\arcsec{\hbox{$^{\prime\prime}$}}
\def\degree{$^{\circ}$} 
\def\mic{$\mu$ }
\def\ang{\AA }  % Angstrom
\def\cm2{cm$^2$ }
\def\se1{s$^{-1}$ }

\def\gtsima{$\; \buildrel > \over \sim \;$}
\def\ltsima{$\; \buildrel < \over \sim \;$}
\def\prosima{$\; \buildrel \propto \over \sim \;$}
\def\gsim{\lower.5ex\hbox{\gtsima}}
\def\lsim{\lower.5ex\hbox{\ltsima}}
\def\simgt{\lower.5ex\hbox{\gtsima}}
\def\simlt{\lower.5ex\hbox{\ltsima}}
\def\simpr{\lower.5ex\hbox{\prosima}}
\def\la{\lsim}
\def\ga{\gsim}
\def\Lsun{\L_\odot}
\def\sr{4U~1708--40}

\def\ie{{i.e.~}}
\def\eg{{\frenchspacing\it e.g. }}
\def\etal{{et al.~}}

%=============================================================================
\title[Large-scale radio jet from GX~339--4]  
{A transient large-scale relativistic radio jet from GX~339--4 } 
\author[Gallo \etal ]
{E. Gallo$^{1}$\thanks{egallo@science.uva.nl},
S. Corbel$^{2}$,
R. P. Fender$^{1}$,
T. J. Maccarone$^{1}$, A. K. Tzioumis$^{3}$
\\ \\
$^{1}$ Astronomical Institute `Anton Pannekoek' and Center for High Energy
Astrophysics, University of Amsterdam, Kruislaan 403, \\
1098 SJ Amsterdam, The Netherlands. \\
$^{2}$ Universit\'e Paris VII and Service d'Astrophysique (F\'ed\'eration
APC), CEA-Saclay, 91191 Gif-sur-Yvette, France \\
$^{3}$ Australia Telescope National Facility, CSIRO, Post Office Box 76,
Epping NSW 1710, Australia} 
\maketitle
%==================================================================================
\begin{abstract} 
We report on the formation and evolution of a large-scale,
synchrotron-emitting jet from the black hole candidate and X-ray binary system
GX~339$-$4. In 2002 May, the source moved from a low/hard to a very high X-ray
state, contemporaneously exhibiting a very bright optically thin radio
flare. Further observations with the Australia Telescope Compact Array have
tracked the formation of a collimated structure extending to about 12 arcsec,
with apparent velocity greater than 0.9$c$. The luminosity of the outflow
seems to be rapidly decreasing; these observations confirm that transient
large-scale jets are likely to be common events triggered by X-ray state
transitions in black hole X-ray binaries.
\end{abstract}   
%==================================================================================
%=======================================================================
\section{Introduction}
%========================================================================
The X-ray binary GX~339$-$4 comprises a compact primary which is a strong
Black Hole Candidate (BHC), with mass function of 5.8$\pm$0.5 \msun~(Hynes
\etal 2003) and a secondary which is likely to be an evolved low mass star
(Shahbaz, Fender \& Charles 2001; Chaty \etal 2002). The system has an orbital
period of 1.75 days (Hynes \etal 2003) and is located at a distance of at
least 4 kpc (Zdziarski \etal 1998; see also Shahbaz, Fender \& Charles 2001
and Maccarone 2003), with an orbital inclination of less than 60\degree~to the
line of sight, as inferred from the lack of eclipses (Cowley \etal
2002). GX~339$-$4 has been a key source in our understanding of the relation
between accretion and the production of relativistic jets. It was the first
BHC to reveal a positive correlation between radio and X-ray fluxes in the
low/hard X-ray state (Hannikainen \etal 1998, later quantified by Corbel \etal
2003), and to demonstrate an association between the `quenching' of core radio
emission and the transition to a high/soft X-ray state (Fender \etal 1999; see
e.g. Done 2001, for a review on X-ray states and Fender 2004 for
a comparison of radio and X-ray behaviour in Galactic BHCs). After spending
almost three years in `quiescence', GX~339$-$4 re-brightened in X-rays at the
end of 2002 March (Smith \etal 2002) and changed rapidly to a soft outburst
state, undergoing a dramatic state change in 2002 May.  This transition was
associated with a bright radio flare (Fender \etal 2002), reaching four to
five times the brightest radio level ever observed from the source (see Corbel
\etal 2000 for the long-term behaviour of GX~339--4). By analogy with other
systems (\eg XTE~J1550--564, Corbel \etal 2001), this flare was likely to be
the signature of a powerful ejection event.  Repeated radio observations of
GX~339$-$4 have indeed confirmed this association: the 2002 radio flaring has
led to the formation of a large-scale relativistic radio jet, whose morphology
and dynamics will be presented in the course of this Letter.
%========================================================================
\section{ATCA observations} 
%========================================================================
The Australia Telescope Compact Array (ATCA) performed eight continuum
observations of GX~339$-$4 at roughly regular intervals between 2002 April and
August, simultaneously at 4800 MHz (6.3 cm) and 8640 MHz (3.5 cm).  Three
further observations were performed in 2003 January, March and May at four
frequencies: 1384 MHz (21.7 cm), 2368 MHz (12.7 cm), 4800 and 8640 MHz.  

\begin{figure}
\psfig{figure=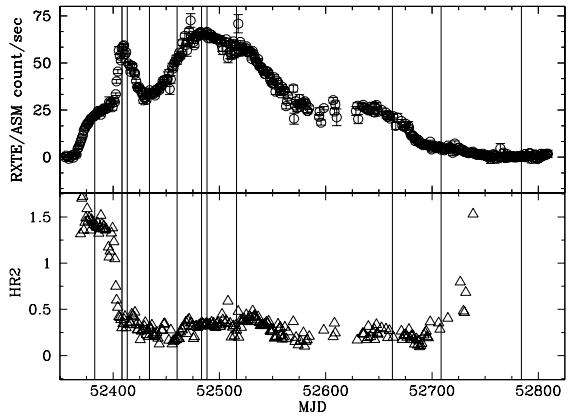,width=8cm,angle=-90}
\caption{\label{fig1}
  X-ray light curve and hardness ratio HR2 (5--12 keV / 3--5 keV count rate,
  only plotted for fractional errors $<0.25$) of GX~339$-$4 as monitored by
  the \it{Rossi X-ray Timing Explorer}\rm/All Sky Monitor.  At the end of 2002
  April, following three years of `quiescence', the source re-brightened in
  X-rays, contemporaneously undergoing a very bright radio flare (on 2002 May
  14, MJD 52408).  The times of our ATCA observations are indicated with solid
  vertical lines; the thick line marks the peak radio level of $\sim$55 mJy. }
\end{figure}

The ATCA synthesis telescope is an east-west array of six 22~m antennas with
baselines ranging from 31~m to 6~km; it uses orthogonal polarized feeds and
records full Stokes parameters (I, Q, U, V).  The target was systematically
offset by about 10 arcsec from the array phase centre, in order to avoid
possible artefacts due to system errors such as DC-offsets. In each
observation, PKS~1934--638 was used for absolute flux and bandpass
calibration, while either PMN~1603--4904, PMN~1650--5044 or PMN~1726--5529 was
the phase calibrator for antenna gains and phases. The
data reduction process and image analysis have been carried out with the
Multichannel Image Reconstruction, Image Analysis and Display (MIRIAD)
software package (Sault, Teuben \& Wright 1995; Sault \& Killeen 1998).  Dates
of ATCA observations are indicated in Fig. \ref{fig1}, superimposed on the
\emph{ Rossi X-ray Timing Explorer}/All Sky Monitor (RXTE/ASM) 2--12 keV light
curve of the system.
%=================================================================
\subsection{Outburst and optically thin core radio flare}
%-----------------------------------------------------------------
%--------------------------------------------------------
Smith \etal (2002) reported an increase in the X-ray flux from GX~339$-$4,
which had been in quiescence for almost three years (Kong \etal 2000), on 2002
March 26 (MJD 52360).  The source reached a peak flux (2--12 keV) of $\sim$0.8
Crab on 2002 May 15 (MJD 52410), followed by a decrease to $\sim$0.4 Crab, and
a secondary rise to $\sim$0.9 Crab around MJD 52500, after which GX~339$-$4
has started a slow return to `quiescence' (reached by the time of
writing). The hardness ratio HR2 suggests that a rapid transition from a
low/hard to a softer X-ray state took place a few days before the (first)
outburst peak, although colour and timing analysis of the RXTE Proportional
Counter Array data indicates a smooth transition from a low/hard to a very
high state in the rising phase, followed by a high/soft state (or possibly an
`oscillating' very-high state) after the peak (Belloni \etal 2002; Nespoli
\etal 2003).

ATCA observations performed between 2002 April and June have detected the
brightest radio flare ever observed by the system, which reached a peak flux
density of about 55 mJy on May 14 (MJD 52408), almost contemporaneously with
the (first) X-ray peak.
%---------------------------------------------------------------------
\begin{table*}
\caption{\label{table1}  
  Image properties of GX~339$-$4 as observed by ATCA at 8640 and 4800 MHz
  between 2002 April and August. Fluxes and positions have been derived by
  fitting the knots with point--like sources; position offsets are expressed
  in arcsec with respect to the binary core position. Flux density errors
  correspond to the rms noise levels in the final, naturally weighted images;
  upper limits are given at a 3$\sigma$ confidence level.  }
\centering
\begin{tabular}{lcccccccc}
\hline
Date	   &MJD	 &8640 MHz & offset~\bf(a)\rm     & lin pol        &  4800 MHz
&offset~\bf(a)\rm& lin pol & $\alpha$\\
(UT time)  &(day)& flux (mJy)    & $\alpha, \delta$ (arcsec)&P/I, PA  &
flux (mJy) & $\alpha, \delta$ (arcsec) & P/I, PA &($S_{\nu}\propto
\nu^{\alpha}$) \\ 
\hline  
2002:04:18  & 52382.75 & 13.49$\pm$0.08 & --0.05, +0.07	& $< 1\%$&
12.97$\pm$0.07&--0.05, +0.09	& $< 1\%$ & +0.07$\pm$0.01	\\ 
2002:05:14  & 52408.48 & \bf 10--40\rm 	&...&$\sim 9\%,-47$\degree$\pm 2$\degree&
\bf  10--55\rm 	&...	&$\sim 4\%, -35$\degree$\pm 3$\degree	&--0.52$\pm$0.01\bf(b)\rm\\  
2002:05:19  & 52413.42 &  12.27$\pm$0.16& --0.12, +0.13  & $< 3\%$&
20.39$\pm$0.23&--0.12, +0.10& $< 2\%$ & --0.86$\pm$0.03\\
2002:06:09  & 52434.28 & $<$ 0.3 &...	&...		& $<$ 0.4	&...&...&...		\\
2002:07:05  & 52460.29 & 10.62$\pm$0.15 & --0.17, --0.05& $\sim 4\%,-55$\degree$\pm 8$\degree &
14.89$\pm$0.19 &--0.22, --0.00	& $\sim 5\%,-21$\degree$\pm 3$\degree & --0.57$\pm$0.03\\
2002:07:28  & 52483.20 & 10.33$\pm$0.18 & --0.25, --0.02& $\sim 5\%,+33$\degree$\pm 6$\degree &
13.84$\pm$0.29 &--0.27, --0.00 & $\sim 5\%,+56$\degree$\pm 5$\degree & --0.50$\pm$0.04\\ 
2002:08:02  & 52488.28 &  9.45$\pm$0.17 & --0.22, --0.08& $< 1\%$ &
11.98$\pm$0.25 &--0.30, --0.03&$< 2\%$ 	&--0.40$\pm$0.05\\
2002:08:30  & 52516.12 &  1.51$\pm$0.13 & --0.63, +0.02 & $< 10\%$ &
2.47$\pm$0.07  &--0.61, +0.20 	& $< 8\%$ & --0.83$\pm$0.14\\
\hline
\end{tabular}
\flushleft
\bf(a)\rm~
absolute positional uncertainty is of 0.26 arcsec between 2002:04:18 and
2002:06:09, 0.35 arcsec between 2002:07:05 and 2002:08:30.
\bf(b)\rm~
Spectral index as measured at the peak value of 55 mJy.  
\end{table*}
%-----------------------------------------------------------------
\begin{figure}
\psfig{figure=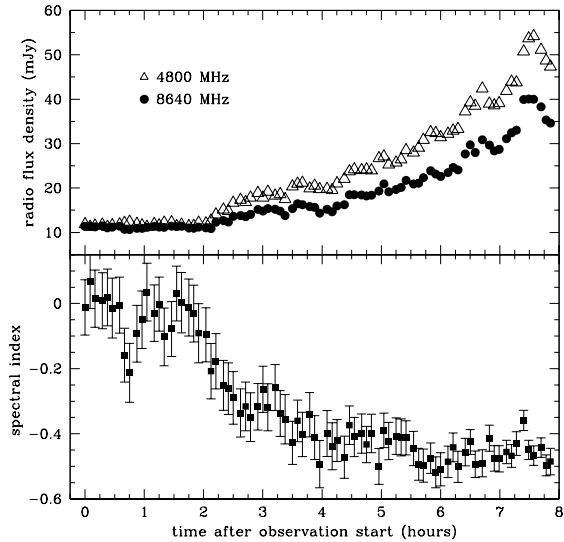,width=7.75cm}
\caption{\label{fig2} 
  Radio light curves (5 minute averages) of GX~339$-$4 on 2002 May 14 (MJD
  52408) at 8640 MHz and 4800 MHz are plotted on the top panel (the typical
  errorbar is smaller than the marker size); the flux density at 4800 MHz rose
  from $\sim$12 to $\sim$55 mJy -- four to five times the brightest radio
  level ever observed from the source -- in 5.5 hours. Temporal evolution of
  the spectral index $\alpha$ (where $ S_{\nu}\propto \nu^{\alpha}$) on the
  bottom panel: as the flux density starts to rise, 2 hours after the
  beginning of the observation, the spectral index starts to decrease. }
\end{figure}
%-----------------------------------------------------------------
The radio flare light curves at 4800 and 8640~MHz are shown in Fig.
\ref{fig2} together with the temporal evolution of the spectral index, which
decreased from $\alpha\sim 0$ down to $\alpha\sim-0.5$ (where
$\alpha=\Delta$log$~ S_{\nu}/\Delta$log$~\nu$). As the
spectral index does not significantly decrease during the last two hours prior
to the flare peak, the flux rise in this time interval can not be due to
decreasing optical depth, as predicted by adiabatic expansion models (\eg van
der Laan 1966), but instead represents a finite phase of particle
acceleration.  

We can derive the minimum energy associated with the emitting component during
the rise, following the formulation by Longair (1994; see also Fender
2004). Assuming an optimal jet viewing angle $\theta\sim26$\degree, for which
the maximum apparent velocity of the jet is achieved (\ie a jet semi-opening
angle given by $cos~\theta=\beta$ with $\beta=0.9$, see next Section), and a
volume of the emitting region of about 9$\times$10$^{44}$cm$^{3}$ (given by
$4/3~\pi\times$($c\times t_{rise})^{3}$, with $t_{rise}= 5.5 $ hours), the
corresponding minimum energy required is $E_{min}~\sim~5\times~10^{39}$ erg;
the associated magnetic field for which the energy in relativistic particles
equals the magnetic energy is of $\sim 8$~mG. The kinetic energy in case of a
pure e$^{+}$e$^{-}$ plasma would be
$E_{kin}~=~(\Gamma~-~1)~\times~E_{min}~\sim~7~\times~ 10^{39}$ erg.~~If there
is one (cold) proton for each electron, then $E_{kin}\sim 5\times10^{40}$ erg,
with an associated mass of $\sim 4\times 10^{19}$g.~~In order to accumulate
such mass for a 10~\msun~BH accreting at a few per cent of the Eddington rate
(as indicated by the X-ray luminosity), it would have taken a few minutes. The
(much longer) observed rise time of 5.5 hours would be of the same order of
the injection time-scale if only a few per cent of the accreted mass was
loaded into the jet.
  
The minimum jet powers equal $3\times 10^{35}$ and $2\times 10^{36}$ erg
s$^{-1}$, for e$^{+}$e$^{-}$ and baryonic plasma respectively.  In the last 20
minutes of the observation, after reaching the peak level of $\sim$~55 mJy at
4800 MHz, the flux density decreases linearly with time with the same slope at
both frequencies, indicating that the main rediative cooling process is
adiabatic (synchrotron and Compton cooling times scale with the
frequency as $\nu^{-0.5}$); the observed decline is much shallower than that
predicted by adiabatic expansion models without any additional energy
injection (\eg $S_{\nu}\propto t^{-4.8}$, van der Laan 1966). \\

The position of the radio source on 2002 April 18, when the source was in a
bright, flat-spectrum radio state prior to the outburst, is consistent with
the binary centre as given by Corbel \etal (2000). Radio emission from
GX~339$-$4 dropped to undetectable levels by 2002 June 09 ($<$0.4 mJy at
4800~MHz), possibly corresponding to extinction of the May 14 flare.

Although further ATCA observations are consistent with a single fading radio
source (see Table~1), observations at 843~MHz performed with the Molonglo
Observatory Synthesis Telescope (MOST) indicate more complex behaviour over
the period 2002 June-July (Campbell-Wilson \& Hunstead, private
communication), possibly associated with multiple ejection events. A more
detailed analysis of the radio variability over this period, including the
MOST data, will be presented in a future paper.

%--------------------------------------------------------------------
\begin{figure}
\hspace{-0.9cm}
\psfig{figure=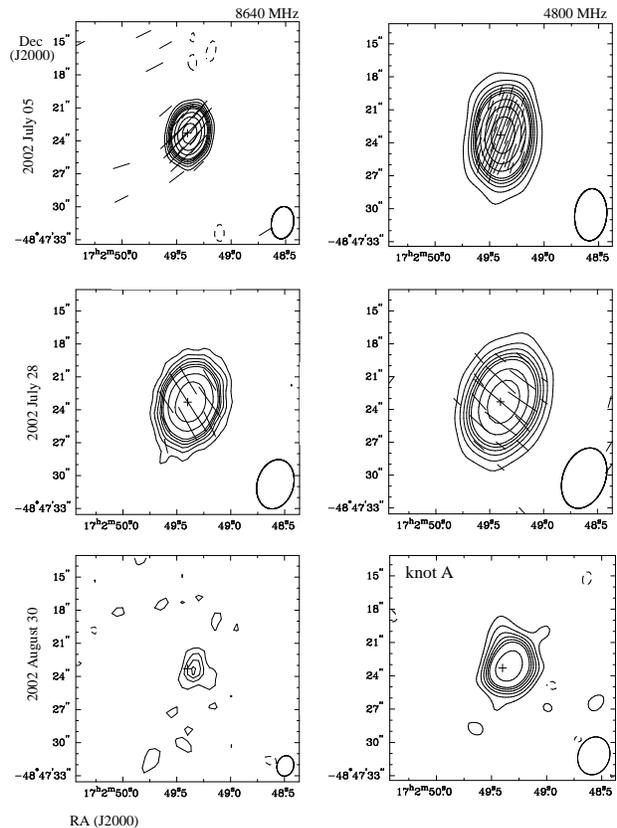,width=9.5cm}
\caption{\label{fig3} 
  Naturally weighted ATCA images of GX~339$-$4 at 8640 and 4800 MHz, left and
  right panels; from top to bottom: 2002 July 05, July 28 and
  August 30; the array configuration was 1.5G, 1.5G and 6C, respectively. The
  contour interval (CI) is chosen as the rms noise in the final image (see
  flux density errors in Table 1), and plotted contours are at --2, 2, 4, 8 ,
  12, 20, 32, 52, 84, 136 times the CI.  Linear polarization E vectors are
  superimposed upon the contour maps in the top and centre panels (where
  significant polarization is detected), showing a clear rotation in the
  position angles. Synthesized beams appear on the bottom right corner of each
  image. The cross indicates the binary core position. By 2002 August 30 the
  radio source is significantly displaced from the core.}
\end{figure}
%--------------------------------------------------------------------------
Fig. \ref{fig3} shows radio maps of GX~339$-$4 at 8640 and 4800 MHz on 2002
July 05, July 28 and August 30. The position of the radio source as imaged by
ATCA is displaced by about 0.2 arcsec to the western side (right on the maps)
of the binary core on 2002 July 05, July 28 and August 02 (not shown),
although, given a total positional uncertainty of 0.3 arcsec in this set of
observations, these coordinates are still consistent with the binary core.  By
2002 August 30, however, the displacement to the western side is of 
$0.6$ arcsec (in position angle PA~=~--~72\degree$\pm$32\degree; PA is
defined positive north--east) at both frequencies, indicating the formation of
a physically separated component. If this event was powered by the May 2002
flare, that would imply a proper motion of about 6 mas/day, that is a minimum
projected velocity of 0.1$c$ (for the minimum distance of 4 kpc); however, if
the displacement of about 0.2 arcsec measured on August 02 was real, it would
imply a separation of at least 400 mas covered in 28 days,
\ie a minimum projected velocity of 0.3$c$.\\
%~~~~~~~~~~~~~~~~~~~~~~~~~~~~~~~~~~`
\subsection{Extended radio jet}
\begin{figure}
\hspace{-1.85cm}
\psfig{figure=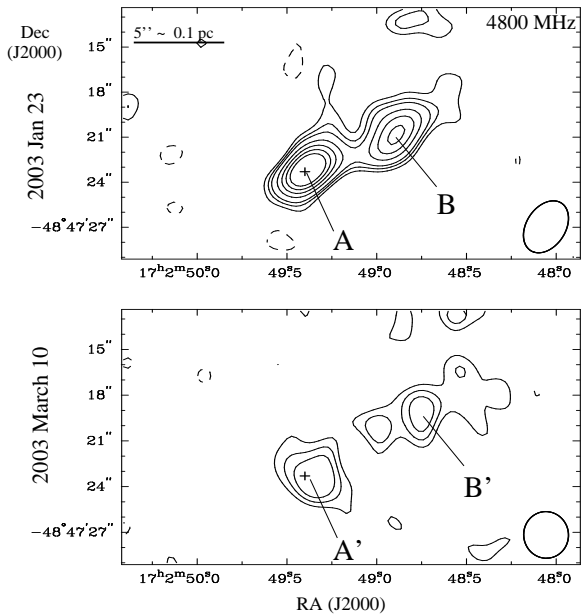,width=12.cm,angle=-90}
\caption{\label{fig4}
  Naturally weighted 4800 MHz maps of the extended jet developed by
  GX~339$-$4.  2003 January 23 is on the top with a peak flux density of 0.66
  mJy/beam and convolved with a Gaussian beam 3.62$\times$2.24 arcsec in
  PA~=~--~22\degree.  Contours are at --2, 2, 4, 6, 8, 10, 12 times the rms
  noise level of 0.04 mJy.  March 10 2003 on the bottom panel, with a peak
  flux density of 0.25 mJy/beam and convolved with a Gaussian beam
  3.06$\times$2.47 arcsec in PA~=~+~1.7\degree. Contours are at --2, 2, 4, 6,
  8, 10, 12 times the rms noise level of 0.02 mJy. Knot B' is displaced by 6.9
  arcsec from the binary core: if powered by the May 2002 flare, this would
  correspond to a projected velocity of about 28~mas/day, that is a jet with
  velocity higher than 0.6$c$ (for D$\simgt$4 kpc).  }
\end{figure}

%~~~~~~~~~~~~~~~~~~~~~~~~~~~~~~~~~~
We expect the bright radio flare(s) of GX~339--4 during its 2002 outburst
to be associated with powerful ejection event(s) fed by the source central
engine. In fact, ATCA has been observing GX~339$-$4 at regular intervals
again in 2003, tracking the formation of a large-scale, relativistic
jet. Fig. \ref{fig4} and Table 2 present the result of these observations at
4800~MHz, where the most notable structures have appeared.  

By 2003 January, an extended outflow composed of two separate `plasmons' has
developed in the same direction of the western component detected on 2002
August 30: the first structure, knot A, is displaced by 0.3 arcsec
north-west from the core, while the other component, knot B, is displaced by
5.5 arcsec in PA~=~--~62\degree~$\pm$2\degree~with respect to the binary core,
implying a minimum velocity of $\sim$0.5$c$ if associated with the 2002 May
flare.  It is worth stressing that this value only represents a lower limit on
the velocity not only because of the lower limit on the distance, but also
because it is likely that the knots were energised by the outflow somewhat
earlier than when they were observed (see Section~3). Both knots display steep
optically thin radio spectrum, with $\alpha_{A}=-0.98\pm0.03 $ and
$\alpha_{B}=-0.96\pm0.08$ (probably either because we are looking above the
cooling break frequency, or because of resolution effects). An elongated
structure is visible at 2368 MHz too.

By 2003 March 10 (MJD 52708) the outflow has faded; at least two components
are distinguishable at 4800 MHz: knot A' and B', probably associated with knot
A and B from January 23. They are displaced respectively by 0.5 and 6.9 arcsec
(with a relative error of 0.3~arcsec), with PA(A')~=~--~83\degree$\pm$26\degree
and PA(B')~=~--~66\degree$\pm$2\degree. The spectrum remains optically thin,
with $\alpha_{A}=-0.98\pm0.10$ (while knot B' is significantly detected at
4800~MHz only).  Assuming again an association with the 2002 May flare, the
separation between knot B' and the binary core (6.9 arcsec) corresponds to a
velocity of 0.6$c$ at 4 kpc.  The jet \emph{head} in the 4800 MHz map is about
12 arcsec away from the core, implying a projected extension of 0.23 pc at 4
kpc and a minimum velocity of 0.9$c$. If GX~339$-$4 was instead at a distance
of at least 5.6 kpc, as estimated from the upper limit on the magnitude of the
secondary star (Shahbaz, Fender \& Charles 2001), the jet would become
\emph{superluminal}, with apparent velocity higher than 1.3$c$; while, given a
minimum distance of 7.6 kpc, as inferred by Maccarone (2003) from the typical
soft-to-hard X-ray state transition luminosity of X-ray binaries, the jet
apparent velocity would be higher than 1.8$c$.
%----------------------------------------------------------------------
%~~~~~~~~~~~~~~~~~~~~~~~~~~~~~~~~~~~~~~~~~~~~~~~~~~~~~~~~~~~~~~~~~

By 2003 May 25 (MJD 52784), the isolated optically thin knots have faded below
detectable levels at all wavelengths.  Instead, a central component has
re-brightened at the binary core position, with peak flux density of 0.9 mJy
at 8640~MHz. Core emission at four frequencies is consistent with an inverted
spectrum ($\alpha=+0.72\pm0.04$), characteristic of an optically thick
synchrotron jet, indicating the source core return to a hard state. Table 3
lists core flux densities (1$\sigma$~upper limits on the optically thin
components of the extended jet are given by their errors, \ie rms noise
levels) at four frequencies. No counter-jet has been detected so far.
%-------------------------------------------------------------------
\begin{table}
\caption{Image properties of the large-scale jet developed by GX~339$-$4 at
  4800 MHz. The absolute positional error, mainly given by the uncertainty on
  the phase calibrator position (here PMN 1603--4904), is of 0.26 arcsec.}
\centering
\begin{tabular}{rcccc}
\hline
Date		& MJD   & 4800~MHz       & offset		&spectral index \\
(UT time)	& (day) & flux (mJy)    & $\alpha,\delta~$(arcsec)& $\alpha$,
($S_{\nu}\propto \nu^{\alpha}$) \\ 
\hline 
2003:01:23	& 52662.67&  	   	  & 			&           	\\
knot A		&         &  0.66$\pm$0.04&  --0.25   +0.22	&$-0.98\pm0.03$\\ 	
knot B		&         &  0.47$\pm$0.04&  --4.88   +2.54	&$-0.96\pm0.08$		\\
2003:03:10      & 52708.56&         	   	 &             	&		\\
knot A'		&         & 0.25$\pm$0.04 &    --0.55, --0.07   &$-0.98\pm0.10$ \\
knot B'  	&         & 0.21$\pm$0.04 &    --6.28, +2.80    &	(a)	\\
%\bf 2003:05:25 & 52784.5&       	 &                           \\
%core        	&       &  0.62$\pm$0.04  &    --0.16, +0.07          \\
\hline
\end{tabular}
\flushleft
(a) knot B' is significantly detected at 4800~MHz only.
\end{table}
%----------------------------------------------------------------------------
\subsection{Linear polarization}
%----------------------------------------------------------------------------
Linearly polarized emission is significantly detected at three epochs: on 2002
May 14, when the powerful flare was detected, the mean polarization level is
of 4$\%$ at 4800 MHz and of 9$\%$ at 8640 MHz, with mean polarization angles:
PA(4800)~=~$-35$\degree$\pm$3\degree and
PA(8640)~=~$-47$~\degree$\pm$2\degree.  On 2002 July 05, about 5$\%$ of the
flux is linearly polarized at both 4800 and 8640 MHz, with electric field
vectors in position angle PA(4800)~=~$-21$~\degree$\pm$3\degree and
PA(8640)~=~$-56\pm$6\degree.  A comparable polarization level is seen on 2002
July 28: position angles have switched to
PA(4800)~=~$+56$~\degree$\pm$5\degree and PA(8640)~=~$+33$\degree$\pm$6\degree
(E vectors are plotted in Fig. \ref{fig3} superimposed upon the contour
maps). The rotations $\Delta$PA(4800~MHz) and $\Delta$PA(8640~MHz) are
consistent between 2002 July 05 and July 28, likely reflecting an overall
rotation of the projected magnetic field of the emission region. The rotation
angle between the two frequencies is between 24\degree--~46\degree on July 05
and between 13\degree--~35\degree on July 28, indicative of foreground Faraday
rotation; if so, we get a lower limit of about 100 rad~m$^{2}$ on the rotation
measure.  No significant linearly polarized emission is detected in any of the
other 2002 observations when the source was bright enough to detect a
polarized signal at a level of a few per cent (see Table 1 for upper
limits). In particular, linear polarization is lower than 1$\%$ by 2002 April
18, when the optically thick core emission is seen ($\alpha$~=~+0.07); for
comparison, Corbel \etal (2000) detected about 2$\%$ of linear polarization at
8704 MHz in 1997 February, with a flux of 8--9 mJy and a similar optically
thick spectrum ($\alpha$~=~0.11--0.23).
%=============================================================================
\section{Summary and discussion}
%============================================================================
The main result established in this work is the formation of a large-scale
relativistic radio jet powered by GX~339$-$4 during its 2002 outburst: this
association indicates that large-scale outflows are likely to be \emph{common}
following any major radio flare triggered by a hard-to-soft(er) X-ray state
transition (see \eg Harmon \etal 1995; Fender \& Kuulkers 2001). 

How does the large-scale jet develop?
The position of the radio source detected by ATCA is consistent with the
binary core until the beginning of 2002 August. Observations between the end
of 2002 August and 2003 March reveal the presence of at least two physically
separated components: the first to appear, always brighter and closer to the
binary core (knot A-A'), has a mean separation of 0.5$\pm$0.1 arcsec to the
western side of the core, while its density flux decreases by a factor 10 in
about 220 days.  The second component -- always fainter -- first appears in
the 2003 maps (knot B-B' in Figure
\ref{fig4}) and is displaced by about 7~arcsec from the binary core in
PA~=~--~64\degree$\pm$2\degree. Once they have appeared, both knots decline in
flux while their positions remain unchanged within uncertainties, unlike \eg
in GRS~1915+105, where the observed ejecta are consistent with simple
ballistic bulk motions (Rodr\'\i guez \& Mirabel 1999; Fender \etal 1999). The
large-scale jet of GX~339-4 seems instead to be better explained in terms of
\emph{shock waves} formed within the jet itself and/or by the 
interaction of an underlying highly relativistic outflow with ambient matter.
A more detailed modelling of the jet components, including maps at all four
frequencies, will be presented in a forthcoming paper.

Polarization analysis at three epochs shows a significant rotation of the
electric vector position angle at two frequencies, indicating a change in the
projection of the magnetic field on the plane of the sky.  The large-scale jet
position angle (PA~=~--~64\degree$\pm$2\degree) is consistent with that of the
jet-like extension in the 8640 MHz map of GX~339$-$4 observed by ATCA in 1996
July (Fender \etal 1997). In addition, Corbel \etal (2000), analysing the
persistent radio emission of GX~339$-$4 while in the low/hard X-ray state,
found a similar PA for the electric field vector (expected to be parallel to
the magnetic field vector in case of optically thick spectrum) of the linearly
polarized signal, indicating a rather stable jet orientation over the years.
%~~~~~~~~~~~~~~~~~~~~~~~~~~~~~~~~~~~~~~~~~~~~~~~~~~~~~~~~~`
\begin{table}
\caption{Core flux densities (mJy) from GX~339--4 on 2003 May 25 (MJD 52784); 
  the spectrum is optically thick, with $\alpha=0.72\pm0.04$.}
\centering
\begin{tabular}{ccccc}
\hline
1384~MHz & 2368~MHz 	& 4800~MHz 	& 8640~MHz\\
\hline
$<0.57$  &$0.39\pm0.08$ & $0.60\pm0.04$	&$0.94\pm0.06$\\
\hline
\end{tabular}
\end{table}
%~~~~~~~~~~~~~~~~~~~~~~~~~~~~~~~~~~~~~~~~~~~~~~~~~~~~~~~~~~`

Persistent large-scale radio jets observed in 1E 1740.7--2942 (Mirabel \etal
1992) and GRS 1758--258 (Rodr\'\i guez, Mirabel \& Mart\'\i~1992) have been
found to extend up to 1--3 pc; in the case of GRS 1915+105 instead,
relativistic ejecta were tracked up a projected distance of 0.08 pc (Mirabel
\& Rodr\'\i guez 1999), while in the large-scale X-ray jet powered by
XTE~J1550--564 (Corbel \etal 2002), the eastern jet has been detected to a
projected physical separation of 0.75 pc covered in 4 years. The large-scale
radio jet of GX~339$-$4 displays something, with a projected extension
of 0.23 pc (at 4 kpc) covered in about 250 days (if associated with the May
2002 radio flare).  Scheduled \emph{Chandra} observation will discover if the
jet is active in X-rays as well, possibly confirming the similarity with
XTE~J1550--564 (Corbel \etal 2002; Kaaret \etal 2003; Tomsick \etal 2003),
where the X-ray jet is still capable of accelerating particles up to TeV
energies four years after the main ejection event.
%=================================================================
\section*{Acknowledgments}
We thank Dick Hunstead and Duncan Campbell-Wilson for kindly providing us
with preliminary results of the MOST observations. The Australia Telescope is
funded by the Commonwealth of Australia for operation as a National Facility
managed by CSIRO. RXTE/ASM results are provided by the ASM/XTE team at MIT.
%=================================================================

\end{document}